\documentclass[twocolumn,aps]{revtex4}

\pdfoutput=1
\usepackage[english]{babel}
\usepackage{amsmath,amssymb,amsxtra,amsfonts}
\usepackage{graphicx}
\usepackage{dcolumn}
\usepackage{txfonts}

\def\bea{\begin{eqnarray}}
\def\eea{\end{eqnarray}}
\def\be{\begin{equation}}
\def\ee{\end{equation}}

\newcommand{\SA}{\texttt{SA} }	

\begin{document}

\title{Physical condition for the slowing down of cosmic acceleration}
\author{Ming-Jian Zhang$^{b}$}
\author{Jun-Qing Xia$^{a}$}\email[Corresponding author: ]{xiajq@bnu.edu.cn}
\affiliation{$^a$Department of Astronomy, Beijing Normal University, Beijing 100875, China}
\affiliation{$^b$Key Laboratory of Particle Astrophysics, Institute of High Energy Physics,
Chinese Academy of Science, P. O. Box 918-3, Beijing 100049, China}

\begin{abstract}
The possible slowing down of cosmic acceleration was widely studied. However, the imposition of dark energy parametrization brought some tensions. In our recent paper, we test this possibility using a model-independent method, Gaussian processes. However, the reason of generating these tensions is still closed. In the present paper, we analyse the derivative of deceleration parameter to solve the problems. The reconstruction of the derivative again suggests that no slowing down of acceleration is presented within 95\% C.L. from current observational data. We then deduce its constraint on dark energy. The corresponding constraint clearly reveals the reason of tension between different models in previous work. We also study the essential reason of why current data cannot convincingly measure the slowing down of acceleration. The constraints indicate that most of current data are not in the allowed region.

\end{abstract}

\maketitle

\section{Introduction}
\label{introduction}

Multiple experiments have consistently confirmed the cosmic late-time accelerating expansion. Contributions to this pioneering discovery contain the type Ia supernova (SNIa) \citep{riess1998supernova,perlmutter1999measurements}, large scale structure \citep{tegmark2004cosmological}, cosmic microwave background (CMB) anisotropies  \citep{spergel2003wmap}, and baryon acoustic oscillation (BAO) peaks \citep{eisenstein2005detection}. One theoretical paradigm to describe the acceleration is the exotic dark energy with repulsive gravity. In the dark energy doctrine, a large number of phenomenological models were invented in terms of equation of state (EoS) $w(z)$. In particular, the Chevallier-Polarski-Linder (CPL) \cite{chevallier2001accelerating,linder2003exploring} model has attracted great attention. As well as the dynamical theory, kinematics is another way to understand the cosmic acceleration. The deceleration parameter $q(z)<0$ is just a direct expression of the phase transition of accelerating expansion. Instead, $q(z)>0$ is a symbol of the decelerating expansion.

Recently, a model of slowing down of acceleration (hereafter \SA)  has caused wide public concern \cite{shafieloo2009cosmic}. In the prior of CPL parametrization, the authors found that the cosmic acceleration may have already peaked and we are currently witnessing its slowing down at $z \lesssim 0.3$. In other words, deceleration parameter $q(z)$ may be changing from negative to positive, or has been achieved a positive value. Following this research, more observational data and dark energy parameterized models were invested in the subsequent investigations \cite{li2010probing,li2011examining,cardenas2012role,2013PhRvD..87d3502L}. Including the two comprehensive studies \cite{magana2014cosmic,2016ApJ...821...60W}, they generally believed that the speculation of  \SA was pendulous. We found that the unconvincing results mainly lie in the two facts that there exists a tension between different models and a tension between different data.  Consequently, a model-independent test is really necessary to better understand the cosmic evolution.

In our recent work \cite{zhang2016test}, we presented a model-independent analysis on this interesting subject, using the powerful Gaussian processes (GP) technique. Unlike the parametrization constraint, this approach does not rely on any artificial dark energy template. It is thus able to faithfully model the cosmology. Using the public code GaPP (Gaussian Processes in Python) invented by \citet{seikel2012reconstruction}, we studied the deceleration parameter with abundant data including luminosity distance from Union2, Union2.1 compilation and gamma-ray burst, and Hubble parameter from cosmic chronometer and baryon acoustic oscillation peaks. The
GP reconstructions suggest that no \SA is detected within 95\% C.L. from
current observational data.

However, above tensions still puzzle us. The reason of why some models can lead to the \SA, while some ones cannot, is still not available. To solve this problem, we deduce the derivative of deceleration parameter which can draw a picture on it. Our goal in the present paper, on the one hand, is to reconstruct the derivative of deceleration parameter by GP method to further test the \SA. On the other hand, we try to provide a physical condition or constraint on the dark energy and observational data, to answer why current data cannot present the \SA.

In Section~\ref{methodology}, we first introduce the methodology including some theoretical basics on the \SA, the GP approach and relevant data. We then present the reconstruction result from current data in Section \ref{result}. The possible physical condition is analyzed in Section \ref{sec:condition} before we discuss conclusions in Section \ref{conclusion}.

\section{Methodology}
\label{methodology}

In this section, we briefly deduce the theoretical formulas for the \SA, and then describe the reconstruction method and observational data used in the present work.

\subsection{Theoretical basics}
\label{basics}

\begin{figure}
    \begin{center}
\includegraphics[width=6cm,height=9cm]{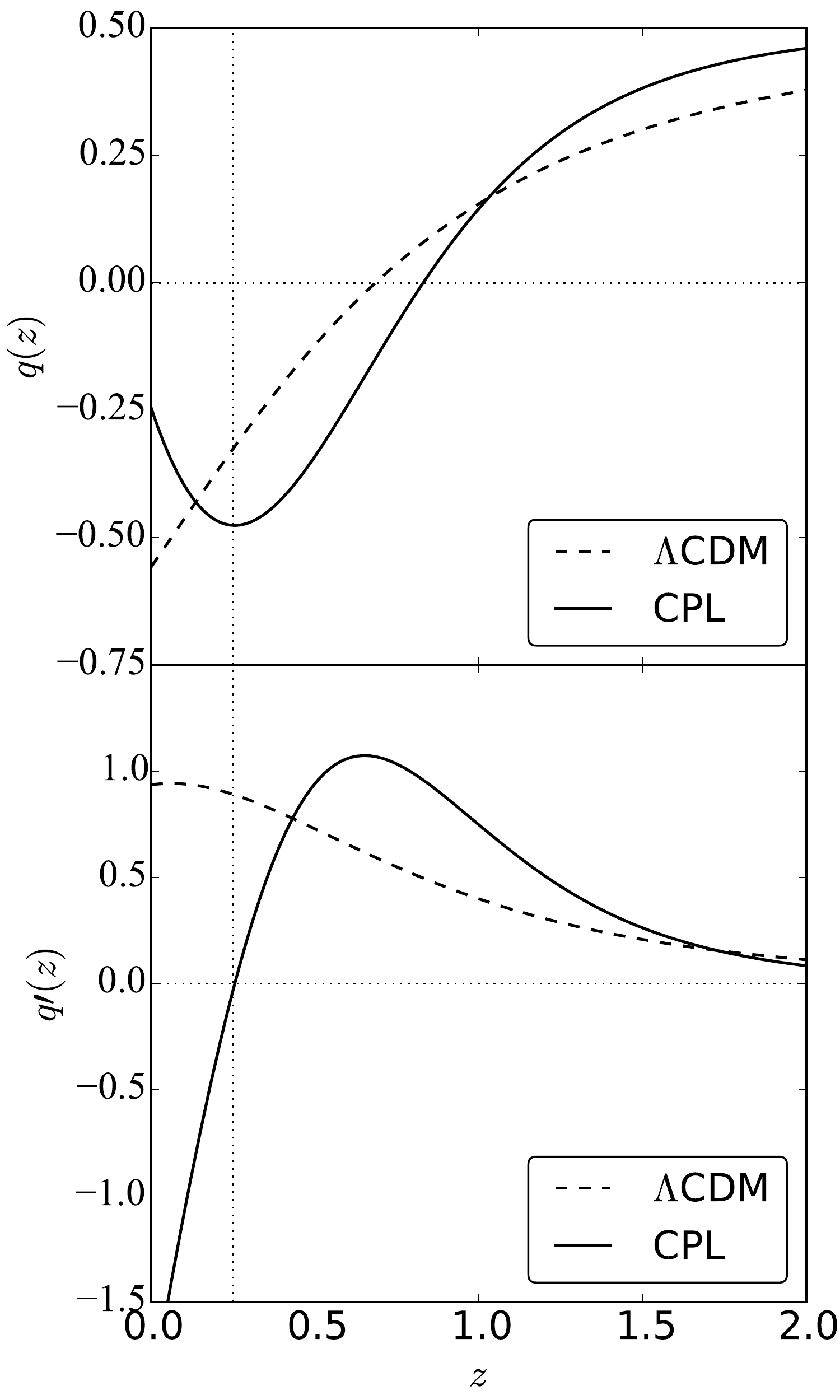}
    \end{center}
    \caption{\label{example} Illustration for the slowing down of cosmic acceleration in the $\Lambda$CDM model with $\Omega_m=0.3$ and CPL model with $w_0=-0.705$ and $w_a=-2.286$ \cite{2016ApJ...821...60W}.}
\end{figure}

In the FRW framework, the distance modulus of SNIa can be estimated as
    \begin{equation}  \label{mu:define}
    \mu (z) = 5 \textrm{log}_{10}d_L(z)+25,
    \end{equation}
where luminosity distance function $d_L(z)$ in the spatial flatness assumption is
\begin{equation}
    \label{dL:define}
    d_L(z) = (1+z)  \int^z_0 \frac{ \mathrm{d} z'}{H(z')} .
\end{equation}
For convenience, we define the dimensionless comoving luminosity distance
\be   \label{D:define}
D(z) \equiv H_0 (1+z)^{-1} d_L (z) ,
\ee
and the dimensionless Hubble parameter
\be   \label{h:define}
h (z) \equiv \frac{H(z)}{H_0} .
\ee
Therefore, substituting Eq. \eqref{D:define} into Eq. \eqref{mu:define}, the distance $D(z)$ from the observational distance modulus is derived. In the same way, we can obtain the dimensionless Hubble parameter via \eqref{h:define}. Certainly, Eqs. \eqref{D:define} and \eqref{dL:define} also contain a potential relation
\be  \label{Hz:define}
h(z) = \frac{1}{D'} ,
\ee
where the prime denotes derivative with respect to redshift $z$. Now the deceleration parameter can be easily expressed as
\begin{equation}  \label{qH:define}
   q(z)=  \frac{h'}{h} (1+z) -1  ,
\end{equation}
and
\be  \label{qD:define}
q(z) = \frac{-D''}{D'} (1+z) -1 .
\ee
In common picture, cosmic evolution can be expressed by the deceleration parameter $q(z)$. Comprehensive work \cite{2016ApJ...821...60W} shows that current deceleration parameter within 95\% C.L. is $q_0 >0$ under the popular CPL parametrization $w=w_0 + w_a z/(1+z)$ with $w_0=-0.705 ^{+0.207}_{-0.212}$ and $w_a=-2.286^{+1.675}_{-1.469}$. However, in our recent work \cite{zhang2016test} we found that $q_0>0$ cannot be favored by the GP method within 95\% C.L. Therefore, to further investigate whether the cosmic acceleration has reached its peak, we need the derivative of deceleration parameter $q'(z)$. This is due to the fact that peak in the $q(z)$ corresponds to an extremum point in its derivative $q'(z)$, as shown in Figure \ref{example}. Using the best-fit values of above CPL model, we calculate that the peak happens at $z=0.25$. Taking derivative on Eq. \eqref{qH:define} and \eqref{qD:define}, we can obtain
\be  \label{dq:Hubble}
q'(z) = \frac{h'}{h} + (1+z) (\frac{h''}{h} - \frac{h'^2}{h^2}) ,
\ee
and
\be  \label{dq:distance}
q'(z) = -\frac{D''}{D'} + (1+z) (\frac{D''^2}{D'^2} - \frac{D'''}{D'}) .
\ee
From the illustration in Figure \ref{example}, we deduce that the \SA would propose a physical condition
\be  \label{physical_condition}
q' (z) \leq 0
\ee
at recent epoch, where the equal sign $q'(z)=0$ corresponds to its peak. In the following, we will use this condition to analyze why some dark energy models and the current data cannot present the \SA.

\subsection{Gaussian processes}  \label{GP}

\begin{figure*}
\includegraphics[width=18cm,height=5cm]{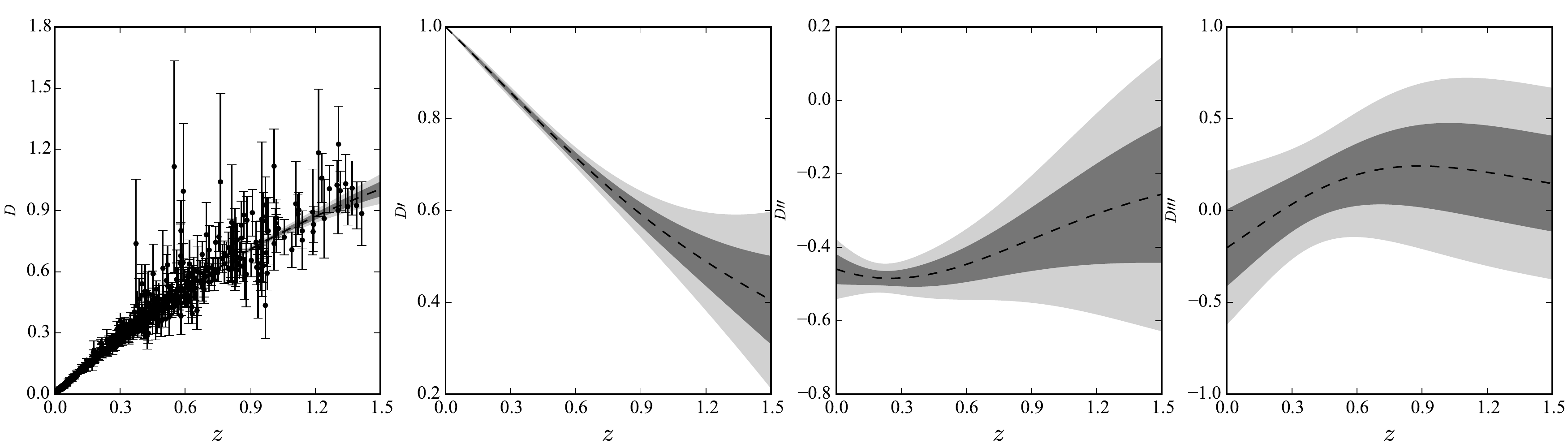}
\caption{\label{GP_union21} The reconstruction of distance $D$ and its derivatives using the Union2.1 supernova data. }
\end{figure*}
In the parametrization constraint, a prior on the constrained function $f(z)$ is imposed, such as the CPL model with two artificial parameters $w_0$ and $w_a$. However, in the GP approach, any parametrization assumption about it is redundant. The key ingredient in this concept is the covariance function $k(z, \tilde{z})$ which connects the function $f(z)$ at different points. Moreover, the covariance function $k(z, \tilde{z})$  only depends on two hyperparameters $\ell$ and $\sigma_f$ which are completely determined by the observational data. Therefore, it has motivated a wide application in the reconstruction of dark energy EoS \cite{seikel2012reconstruction,holsclaw2010nonparametric1,holsclaw2010nonparametric2}, and in test of the concordance model \cite{seikel2012using,yahya2014null}.

For the covariance function $k(z, \tilde{z})$, many templates are available. The usual choice is the squared exponential $k(z, \tilde{z}) = \sigma_f^2 \exp[-|z-\tilde{z}|^2 / (2 \ell^2)]$.  Analysis in Ref. \cite{seikel2013optimising} shows that the Mat\'ern ($\nu=9/2$) mode is a better choice to present suitable and stable result. It thus has been widely used in previous work \cite{yahya2014null,yang2015reconstructing}. It is read as
\begin{eqnarray}
k(z,\tilde{z}) &=& \sigma_f^2
  \exp\Big(-\frac{3\,|z-\tilde{z}|}{\ell}\Big) \nonumber \\
  &&~\times \Big[1 +
  \frac{3\,|z-\tilde{z}|}{\ell} + \frac{27(z-\tilde{z})^2}{7\ell^2}  \nonumber\\
&&~~~~~~
+ \frac{18\,|z-\tilde{z}|^3}{7\ell^3} +
  \frac{27(z-\tilde{z})^4}{35\ell^4} \Big]. \label{mat}
\end{eqnarray}
With the chosen Mat\'ern ($\nu=9/2$) covariance function, we can reconstruct the derivative $q' (z)$ using the publicly available package GaPP \cite{seikel2012reconstruction}. It also has been frequently used in above referenced work.

\subsection{Observational data}  \label{data}

The dataset we use here are the supernova data from Union2.1 compilations and the observational $H(z)$ data (OHD). For the Union2.1 data, they were released by the Hubble Space Telescope Supernova Cosmology Project, which contains 580 samples on the distance modulus with errors \cite{suzuki2012hubble}. In this family, it can span the redshift region up to $z<1.414$, but most of them are in the low redshift region, which is helpful to present a clear understanding on the \SA at recent epoch.
Following previous work \cite{seikel2012reconstruction,yahya2014null,yang2015reconstructing}, we can fix $H_0=70$ km s$^{-1}$Mpc$^{-1}$ and include the covariance matrix with systematic errors. To obtain the distance $D(z)$, we should convert the distance modulus into it via Eq. \eqref{D:define}. Moreover, the theoretical initial conditions $D(z=0)=0$ and $D'(z=0)=1$ should be taken into account in the calculation.

For the OHD, two ways are available to get them. One is from the differential ages of galaxies
\cite{jimenez2008constraining,simon2005constraints,stern2010cosmic}, usually called cosmic chronometer. The other is
from the BAO peaks in the galaxy power spectrum
\cite{gaztanaga2009clustering,moresco2012improved} or from the BAO
peak using the Ly$\alpha$ forest of QSOs \cite{delubac2013baryon}. In this paper, we use the most recent dataset in Table I of our previous work \cite{zhang2016test}. This catalog accommodates 40 data points, which includes the latest five measurements by \citet{moresco2016A}. After the preparation of OHD, we should normalize the $H(z)$ to generate the dimensionless $h(z)=H(z)/H_0$. The initial condition encoded is $h(z=0)=1$. Considering the error of Hubble constant, we can obtain the uncertainty of $h(z)$
\be
\sigma_h^2 = \frac{\sigma_H^2}{H_0^2} + \frac{H^2}{H_0^4} \sigma_{H_0}^2 .
\ee
In our calculation, we utilize the prior from recent measurement $H_0=73.24 \pm 1.74$ km s$^{-1}$ Mpc$^{-1}$ with 2.4\% uncertainty by the Hubble Space Telescope \cite{2016ApJ...826...56R}.

\begin{figure}
\centering
\includegraphics[width=0.35\textwidth]{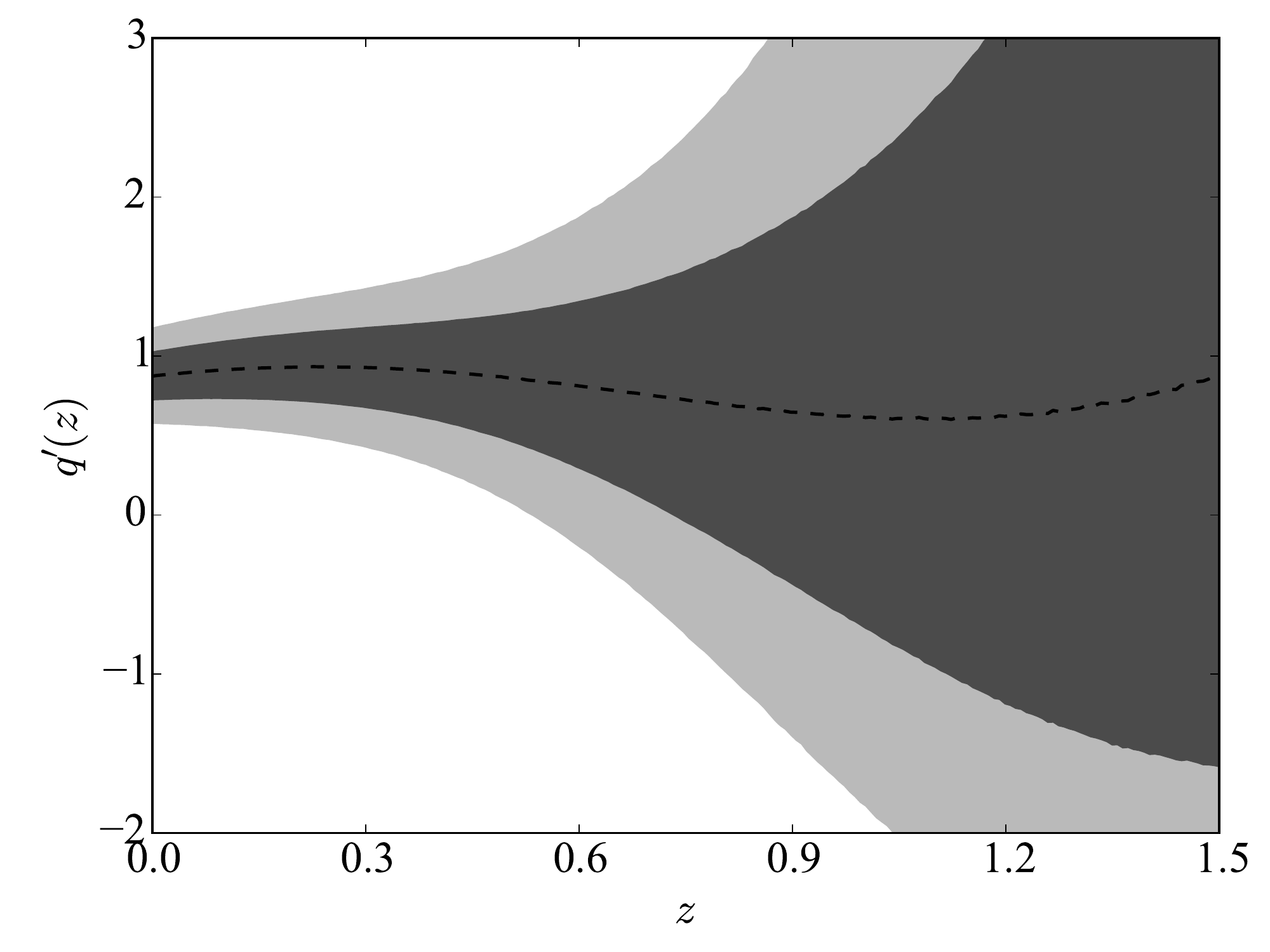}
\caption{\label{dq_union21} The reconstruction of $q'(z)$ using the Union2.1 supernova data.}
\end{figure}

\section{Result}
\label{result}

\begin{figure*}
\includegraphics[width=18cm,height=5cm]{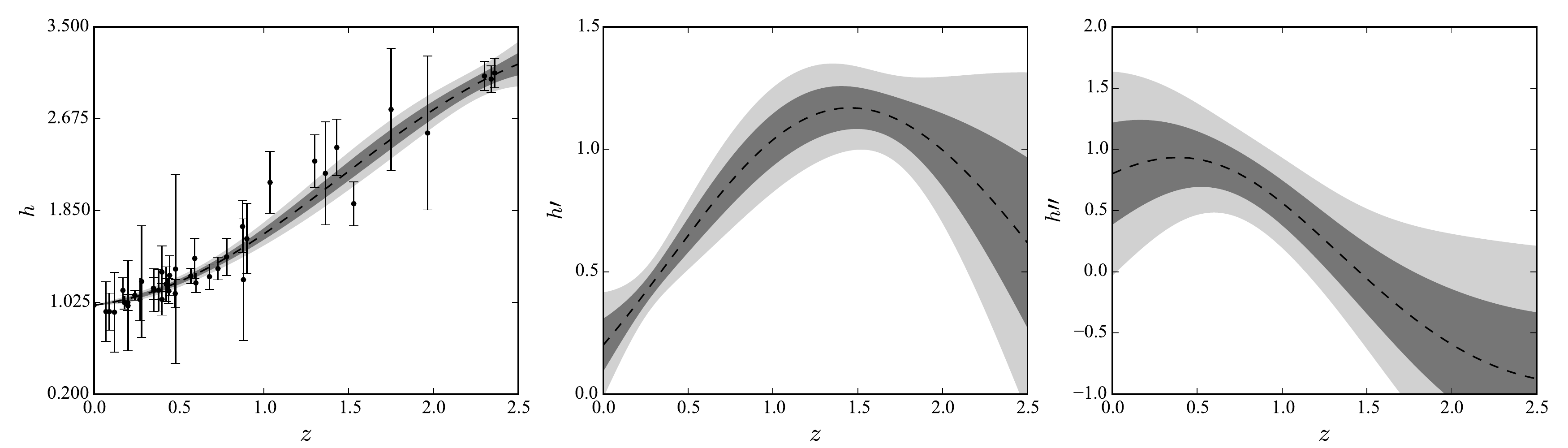}
\caption{\label{GP_OHD} The reconstruction of Hubble parameter $h$ and its derivatives using the OHD.}
\end{figure*}

\subsection{Reconstruction from the Union2.1 data}
\label{resultUnion21}

We now report the GP reconstruction for the Union2.1 SNIa data, as shown in Figs. \ref{GP_union21} and \ref{dq_union21}. The dashed lines correspond to the mean values of reconstruction. Shaded regions are reconstruction with 68\% and 95\% C.L.

The first panel of Fig. \ref{GP_union21} shows that distance $D$ agrees very well with the observational data, which indicates that the GP is reliable. While for the derivative $D'$ and $D''$, the former is positive and decreases sharply; the latter increases slowly with the redshift. Moreover, $D''$ in most redshift region is negative with a current estimation $D''(z=0)=-0.45 \pm 0.04$. We also note that errors of all the reconstructions increase with the redshift, especially for $z \gtrsim 0.6$. This is because GP approach is sensitive to the observational error. Uncertainties at high redshift are bigger than those at low redshift. Moreover, number of the high-$z$ supernova data is smaller than that of the low-redshift. Therefore, improvement of measurement precision at high redshift is very necessary.

In Fig. \ref{dq_union21}, we plot the $q'(z)$ reconstruction. First, mean value shows that $q'(z)$ oscillates stably around the value of one. Second, but the most importantly, we obtain that $q'(z)$ within 95\% C.L. at recent epoch is positive. The reconstruction presents that current supernova data do not favor the \SA  at recent period, even a trend of the decreasing $q'(z)$.

\begin{figure}
\centering
\includegraphics[width=0.35\textwidth]{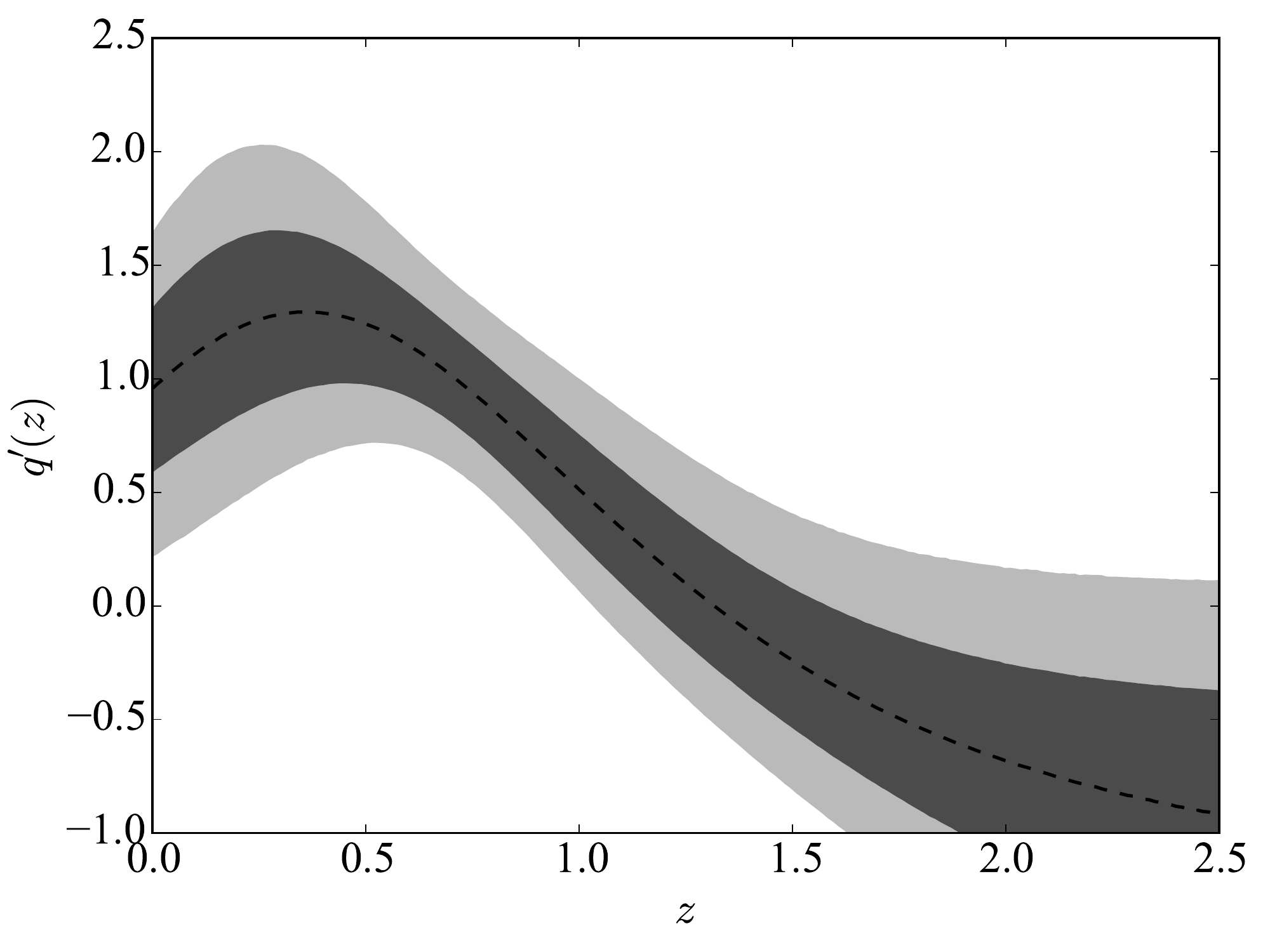}
\caption{\label{dq_OHD}  The reconstruction of $q'(z)$ using the OHD.}
\end{figure}

\subsection{Reconstruction from the OHD}
\label{resultOHD}

Fig. \ref{GP_OHD} plots the reconstruction of $h$ and its derivatives. First, we note that distribution of OHD is quite different from the supernova data. For the Union2.1 data, they distribute uniformly at low redshift. Moreover, their errors at low redshift are comparatively small. But for the OHD at low redshift, their distribution is not only dispersive, but also possess big errors. This profile would influence the reconstruction inevitably, as shown in other two panels. Second, we find that values of $h'$ and $h''$ both are positive at low redshift within 95\% C.L. Specifically, it gives $h'(z=0)=0.2 \pm 0.1$.

From the $q'(z)$ reconstruction in Fig. \ref{dq_OHD}, we find that at recent period $q' >0$ within 95\% C.L. That is, no evidence hints that the cosmic acceleration has reached its peak. Interestingly, comparison between Fig. \ref{dq_union21} and \ref{dq_OHD} shows that these two different dataset produce a similar estimation $q'(z=0)=0.9$. But different from the supernova data, $q'$ for the OHD at recent epoch is decreasing. Moreover, this trend indicates that $q'$ in the future may be across the zero, leading to the \SA.

\section{Physical condition for the slowing down of acceleration}
\label{sec:condition}

Above reconstructions both imply that current data disfavor the cosmic acceleration has reached its peak at recent period within 95\% C.L. Many previous work found that some dark energy parametrizations, such as the CPL model, led to the \SA. While in the $\Lambda$CDM, this strange phenomenon disappeared. Nevertheless, the reason of this tension was not available. In this section, on the one hand, we try to find out this reason. On the other hand, we will investigate which physical condition should be satisfied for observational data to obtain the \SA.

\subsection{Requirement for the dark energy}
\label{Requirement for the dark energy}

Different from previous work, here we consider the dark energy without any parametrization
\be  \label{DE_model}
h^2(z) = \Omega_m (1+z)^3 + (1-\Omega_m) \exp \left[3 \int^z_0 \frac{1+w(z')}{1+z'} dz' \right] .
\ee
This free form allows us to gain a more objective understanding of the \SA. Next, we investigate which physical condition may be imposed by the \SA on dark energy.

Substituting Eq. \eqref{DE_model} into \eqref{dq:Hubble}, we can obtain the derivative
\be
q' (z) = \frac{1}{2h^4} \left[ h^{2'} h^{2} + (1+z) h^{2''} h^{2} - (1+z) (h^{2'})^{2} \right].
\ee
According to the physical condition \eqref{physical_condition},
we eventually obtain
\be  \label{w_diff}
w' (z) \leq -\frac{3\Omega_m (1+z)^2}{h^2(z)} w^2 .
\ee
Obviously, right of the Eq. \eqref{w_diff} is negative. That is, the \SA at least requires $w' <0$. Therefore, it is not difficult to understand why dark energy model with constant EoS, such as $\Lambda$CDM model and $w$CDM, cannot produce the \SA, no matter what kind of observational data were used. In the same way, dark energy model with a positive $w'$ also cannot generate this strange and interesting phenomenon. Therefore, our above analysis reveals the reason of tension in previous work.

\begin{figure}
\centering
\includegraphics[width=0.35\textwidth]{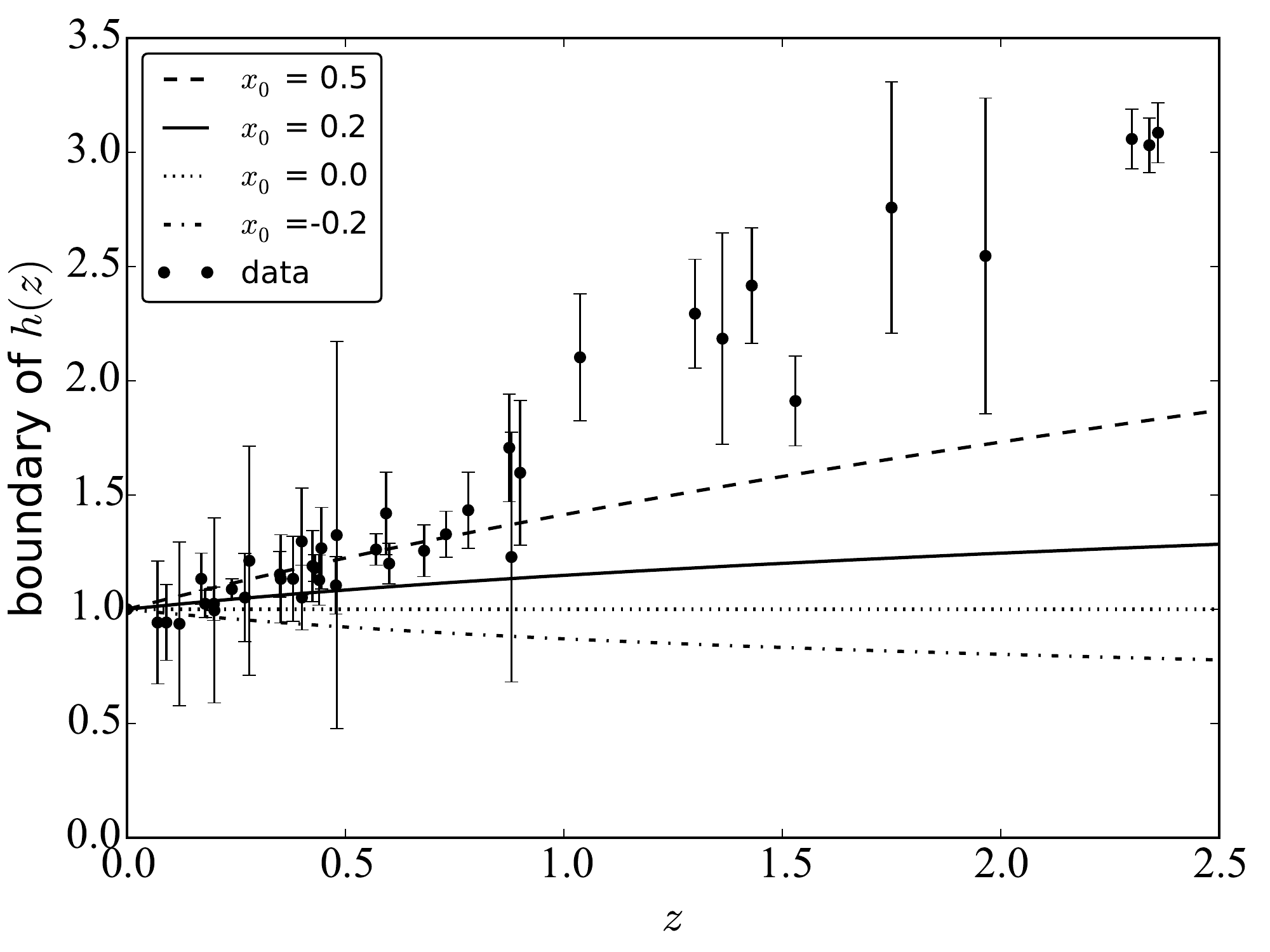}
\caption{\label{comparison1}  Comparison between the boundary of $h$ and OHD. The boundary of $h$ is plotted based on the physical conditions \eqref{h_bound} for different initial value $x_0$. }
\end{figure}

\subsection{Requirement for the observational data}
\label{Requirement for the observational data}

Investigations of the \SA in previous work have not given a clear and consistent estimation. A large part of reason lies on its dependence of dark energy model. Putting aside the dark energy, what kinds of data are required by the \SA?

Let us define a new variable $x \equiv \frac{h'}{h}$, with an initial condition $x_0=h'(z=0)$. Theoretically, it is difficult for us to determine it without any model assumption. The GP reconstruction in Figure \ref{GP_OHD} shows that current observational data give a positive value $h'(z=0)=0.2 \pm 0.1$. However, in order to present a reasonable analysis, we temporarily set the initial condition to be a free parameter. Now, using the new variable $x$, Eq. \eqref{dq:Hubble} consequently can be reduced to an easy form
\be
q' (z) = x + (1+z) x' .
\ee
Through the physical condition \eqref{physical_condition}, we obtain
\be  \label{x_diff}
x' \leq -\frac{x}{1+z} .
\ee
According to the sign of variables $x$ and $x_0$, we can solve this formula in four scenarios. According to the integral comparison theorem, the formula \eqref{x_diff} changes into an inequality of the integral. After a series of calculation, it leads to
\be  \label{h_bound}
h \leq (1+z)^{x_0} ,
\ee
with $x, \; x_0>0$ or $x, \; x_0 <0$. For other cases, no real solution can be obtained, because of the existence of singularity at $x=0$.
So far, we have deduced that an upper boundary \eqref{h_bound} should be satisfied by the OHD to obtain the \SA.

\begin{figure}
\centering
\includegraphics[width=0.35\textwidth]{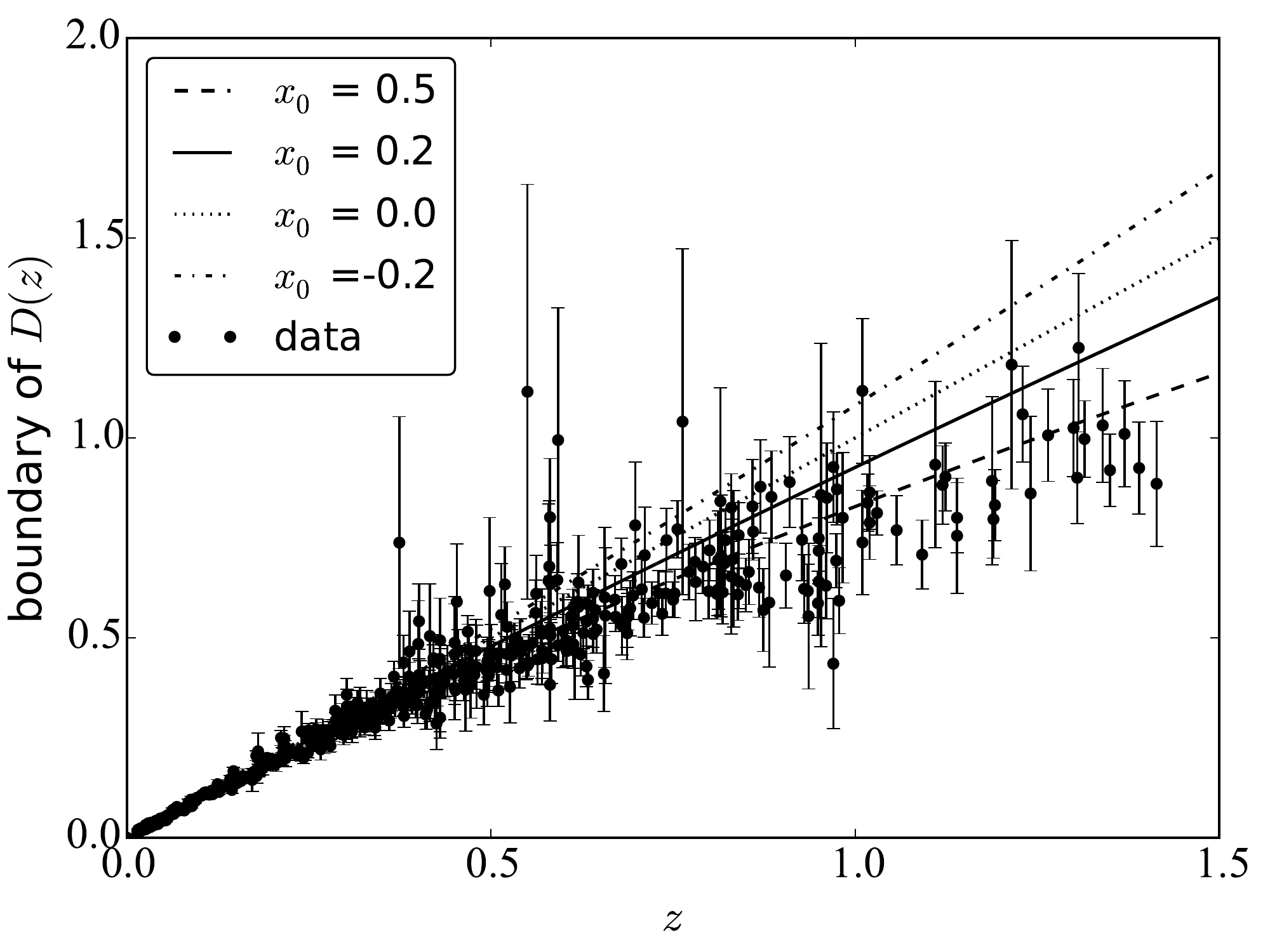}
\caption{\label{comparison2}  Comparison between the boundary of $D$ and Union2.1 data. The boundary of $D$ is plotted based on the physical condition \eqref{D_bound} for different initial value $x_0$. }
\end{figure}
In Fig. \ref{comparison1}, we perform a comparison between the OHD and the boundary of $h$. The lines are the boundary of $h$ at different initial value $x_0$. The points with errorbars are observational data. According to the constraint in Eq. \eqref{h_bound}, region below the lines is allowed region. We find that only few data are in the allowed region. Most of them are active in the illegal region. Therefore,  it is not difficult for us to understand why current OHD cannot produce the \SA . If we want to observe this phenomenon, more data below the constraint line are needed.

Now using the relation \eqref{Hz:define} between $h$ and $D$,  the formula \eqref{h_bound} eventually reduces to
\be
D' \geq  (1+z)^{-x_0}
\ee
for $x, \; x_0>0$ or $x, \; x_0 <0$.
Finally, a constraint of the distance $D$ is given as
\be  \label{D_bound}
D \geq \frac{1}{1-x_0} \big[ (1+z)^{1-x_0} -1 \big] .
\ee
We should note that $x_0$ here is the initial value of $h'$, not the initial value of $D''$.
In Fig. \ref{comparison2}, we plot the boundary of $D$ with different initial value $x_0$ and compare them with the observational data.
Region above the lines are allowed region by the \SA. Comparison shows that most of the data are not in the allowed region, especially for the high redshift. Thus, it is also not strange that current supernova data do not favor the \SA.

\section{Conclusion and discussion}
\label{conclusion}

Possible slowing down of cosmic acceleration has attracted great attention. However, most investigations were model-dependent. In our recent work \cite{zhang2016test}, we made a model-independent test. However, the reason of why some dark energy models and current data cannot present the \SA is still not available. Our study of the derivative of deceleration parameter $q'(z)$ in the present paper can provide an essential analysis and solution to the problems. We actualize this plan via playing a constraint on the dark energy and observational data. 

One difference of our present work from previous investigations was our development of the parameter space. Previously, studies on the \SA were in the deceleration parameter space. In the present paper, we caught the feature of evolution of $q(z)$, and provided a systematic  analysis covering a general survey of the derivative $q'(z)$. Using the study of $q'(z)$, it is not only able to promote the test of \SA in more detail, but solve the above tensions.

Although many parameterizations were considered in previous work, they are model-dependent. As a result, tensions from different models \cite{li2010probing,li2011examining} or different observational datasets \cite{magana2014cosmic,2016ApJ...821...60W} made this study into a more difficult mode. In contrast, the novel GP can break this limitation and produce more accurate and objective estimation. In our previous work \cite{zhang2016test}, we even took into account the influence of spatial curvature and Hubble constant. They all demonstrated the stability of GP method. In previous work \cite{seikel2012reconstruction,holsclaw2010nonparametric1,holsclaw2010nonparametric2,seikel2012using,yahya2014null}, they also supported the superiority of GP technique.

Since the first study of the \SA in Ref. \cite{shafieloo2009cosmic}, so many work were launched. However, tension between different models has been puzzling us. It was found repeatedly that the $\Lambda$CDM model and $w$CDM cannot reduce to the \SA. Instead, the popular CPL in most cases can lead to the \SA. Nevertheless, the reason of generating this tension was still closed. Our analysis in Section \ref{Requirement for the dark energy} provide a clear explanation on this problem. We found that only the dark energy with EoS in Eq. \eqref{w_diff} can tempt the \SA. No matter which observational data were used. 

Another important contribution to the \SA study is our constraint on observational data. Our presentation of the constraint on distance $D$ and Hubble parameter $h$ is not only model-independent, but reveals the essential reason of why the \SA cannot be observed convincingly by current data. Comparisons between the boundaries and observational data indicates that the data are not in the allowed region, which facilitates our understanding of \SA. Moreover, it also provides a reliable scientific reference for future observation.


\section*{Acknowledgments}
J.-Q. Xia is supported by the National Youth Thousand Talents Program and the National Science Foundation of China under grant No. 11422323. M.-J. Zhang is funded by China Postdoctoral Science Foundation under grant No. 2015M581173. The research is also supported by the Strategic Priority Research Program ¡°The Emergence of Cosmological Structures¡± of the Chinese Academy of Sciences, grant No. XDB09000000.

\bibliography{condition}
\end{document}